# Capturing Distribution Grid-Integrated Solar Variability and Uncertainty Using Microgrids

Alireza Majzoobi, Amin Khodaei
Dept. of Electrical and Computer Engineering
University of Denver
Denver, CO, USA
alireza.majzoobi@du.edu, amin.khodaei@du.edu

Shay Bahramirad
Distribution Planning and Smart Grid
Commonwealth Edison
Oakbrook, IL, USA
shay.bahramirad@comed.com

*Abstract*— The variable nature of the solar generation and the inherent uncertainty in solar generation forecasts are two challenging issues for utility grids, especially as the distribution grid integrated solar generation proliferates. This paper offers to utilize microgrids as local solutions for mitigating these negative drawbacks and helping the utility grid in hosting a higher penetration of solar generation. A microgrid optimal scheduling model based on robust optimization is developed to capture solar generation variability and uncertainty. Numerical simulations on a test feeder indicate the effectiveness of the proposed model.

## Nomenclature

*Sets and Indices:*
- $c$    Superscript for distribution network consumers/prosumers.
- ch/dch    Superscript for energy storage charging/discharging.
- $d/i/j$    Index for loads/DERs/prosumers at the distribution network.
- D/G/S    Set of adjustable loads/dispatchable units/energy storage
- N    Set of consumers and prosumers.
- P/U    Set of primal variables/uncertain parameters.
- $t$    Index for time periods (hour).
- $u$    Superscript for the utility grid.
- ^    Index for calculated variables.

*Parameters:*
- DR/UR    Ramp down/up rate.
- DT/UT    Minimum down/up time.
- E    Load total required energy.
- F(.)    Generation cost.
- MC/MD    Minimum charging/discharging time.
- MU    Minimum operating time.
- $\alpha, \beta$    Specified start and end times of adjustable loads.
- $\rho^M$    Market price.
- $\eta$    Energy storage efficiency.

*Variables:*
- C    Energy storage available (stored) energy.
- D    Load demand.
- I    Commitment state of dispatchable units.
- P    DER output power.
- $P^M$    Utility grid power exchange with the microgrid.
- SD/SU    Shut down/startup cost.
- $T^{ch}/T^{dch}$    Number of successive charging/discharging hours.
- $T^{on}/T^{off}$    Number of successive ON/OFF hours.
- $u$    Energy storage discharging state (discharging: 1, charging: 0).
- $v$    Energy storage charging state (charging: 1, discharging: 0).
- $z$    Adjustable load state (operating: 1, otherwise: 0).
- $\lambda, \mu, \pi$    Dual variables.
- $\alpha$    Reflected operation cost in the master problem.

## I. Introduction

THE global concern for increased greenhouse gas emissions, high hydrocarbon-based fuel costs, and the need for sustainable power generation resources have resulted in a rapid worldwide deployment of renewable energy resources that produce inexpensive and emission-free energy. Solar energy resources are anticipated to be core components of future power systems and viable enablers of sustainable power generation. Solar energy is abundant and pollution free, and despite conventional power generation resources, which normally must be installed close to the source of the fuel, solar energy resources can be installed nearly everywhere the sun shines; hence they significantly facilitate on-site and distributed power generation [1],[2]. Moreover, ease of installation and high efficiency has made the solar energy as the first choice of consumers who are willing to utilize an energy resource to offset a portion of their consumption or are seeking economic benefits from a locally generated power [3],[4]. These advantages have established solar energy as a promising technology that will play a major role in future power systems. Recent drops in solar technology prices, i.e., more than 50% over the last five years, have made this technology more attractive than ever. In 2015, the worldwide installed solar energy capacity grew by 28% to reach a total of 227 GW [5]. In 2016, more than 7 GW solar PV were installed in the U.S. while residential PV was the greatest sector with over 2 GW of installation [6].

In the context of solar power forecasting, [7]-[9] discuss the available methods in forecasting. Accurately forecasting solar power is a major area of research in solar industry and many efforts have been devoted to this inquiry. Solar power forecasting is conducted by simulation, statistical methods, or a combination of the two. All forecasting methods suffer from a level of inaccuracy, which is a result of inherent variability in solar irradiance. Constant fluctuations in solar power make accurate forecasting a formidable task. Even if the hours of volatility and intermittency are identified, it would be very difficult to forecast the exact value of solar power, hence solar forecasts are usually involved with high forecast errors and uncertainty. The two major solar power forecast characteristics are defined as variability and uncertainty. Variability denotes the unavailability of solar power during the day and the fact that solar power is not always available (i.e., intermittency), and the fluctuating nature of solar power in different time scales from seconds to minutes to hours (i.e., volatility). Uncertainty denotes the inability to accurately predict in advance the timing and the magnitude of solar generation variability. Thus, despite all advantages, extensive application of solar power is troublesome for electric utilities due to its variable nature which would require highly flexible power systems to facilitate the integration [10]-[12]. Solar power requires backup generation to compensate for the lost power when the forecasted solar power is not materialized or the forecast is inaccurate [13]-[15]. Therefore, effective coordination of solar power with the existing power system infrastructure in order to achieve a smooth and controllable output is critical to an efficient and sustainable deployment of these resources. The ultimate goal in a coordination scheme is to smooth out the variability of solar power when coordinated and operated with other system components. Common methods are to install and commit additional fast response generation units (to provide additional reserve) and/or to coordinate the solar power with an energy storage system [16], [17]. These methods, however, require installation of capital-intensive resources and typically not very desirable for system operators.

A novel and viable method for addressing the aforementioned challenges and mitigating the negative impact of increasing renewable energy generation on power systems is to leverage available flexibility of existing microgrids in distribution networks. The effectiveness of this method in reducing system ramping is validated [18]-[19], while capturing solar generation uncertainty via distributed microgrids is still an untapped topic. Microgrids, as small and self-sufficient localized power systems, have attracted significant attention in recent years as a result of offered reliability, resilience, power quality, and energy efficiency benefits [20]-[23]. Microgrids' deployment of controllable resources provides a fast-response generation/load mix that can be used for capturing the variability and uncertainty of solar generation. This paper will discuss and model this problem via developing upgraded microgrid optimal scheduling models. The model could also be applied to an off-grid system with storage in place of the utility, which will be carried out in a follow up research.

The rest of the paper is organized as follows. The outline of the proposed microgrid optimal scheduling model is presented in Section II. Section III formulates the problem including problem objective, operation constraints, and the proposed decomposition framework required for uncertainty integration. Section IV presents numerical examples to show the effectiveness of the proposed model when applied to a test distribution feeder. Conclusions are presented in Section V.

## II. Model Outline

Consider a typical distribution feeder consisting of a microgrid and several prosumers (i.e., consumers with ability of producing electricity). The utility grid needs to supply a power of $P_t^u = P_t^M + \sum_{j \in N} P_{jt}^c$ to this distribution feeder. It is assumed that prosumers are equipped with distributed solar generation resources, mainly in the form of rooftop solar, and thus can potentially cause considerable variability and uncertainty in the power that is to be supplied by the utility grid. The microgrid can locally capture the variability and uncertainty of solar generation of these prosumers, so the power that is to be supplied by the utility grid is smoothed out and also will be certain to an acceptable level. The main challenge here is that the solar generation is uncontrollable, and furthermore, prosumers attempt to maximize their economic benefits by partially supplying their local loads using solar generation, or selling back to the utility grid, whenever possible. The microgrid, on the other hand, can be incentivized by the utility grid to capture these variabilities.

The proposed microgrid optimal scheduling problem seeks to minimize the microgrid operation cost, including the local generation cost and the cost of energy purchase from the utility grid, while adjusting its output power in a way that solar generation variability and uncertainty is captured. The problem is developed as a robust optimization model and further decomposed to a master problem (scheduling) and a subproblem (optimal operation) using Benders decomposition. The master problem determines the optimal schedule of available resources, including dispatchable units, energy storage, and adjustable loads by minimizing the local generation cost. The obtained schedule, which includes binary scheduling variables, will be sent to the subproblem. The subproblem finds the worst-case optimal operation solution by minimizing the cost of energy purchase from the utility grid and further determines the optimal dispatch of resources along with the power exchange with the utility grid. As enabled by this decomposition, binary variables are only considered in the master problem which is a mixed-integer programming (MIP) problem, while in the subproblem binary variables are fixed, so the subproblem represents a linear programming (LP) problem. The worst-case solution of the subproblem (which is represented in terms of max-min under the employed robust optimization method) can be easily converted to a maximization problem using duality theory. A polyhedral uncertainty set is considered for modeling the solar generation uncertainty, i.e., the uncertain distributed solar generation is assumed to vary in an interval around the forecasted value. To limit the solution conservatism, i.e., to control the robustness, the number of uncertain parameters that can take their worst-case values are limited using a budget of uncertainty. The budget of uncertainty represents the maximum number of hours per day in which uncertain parameters can deviate from their forecasted value. For example, in the case of a 6-hour budget of uncertainty, uncertain parameter (e.g., solar generation) is allowed a maximum of 6 hours to be at its bounds (either upper or lower) and accordingly the forecasted value will be used for solar generation in the remaining hours. Fig. 1 shows the flowchart of the proposed model.

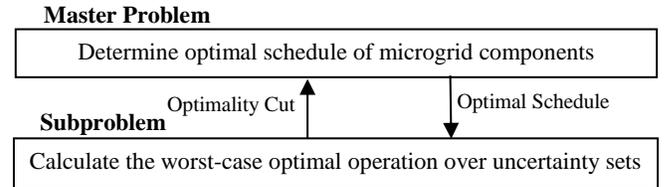

Figure 1. Flowchart of the proposed microgrid optimal scheduling model.

## III. Problem Formulation

### A. Master problem (Scheduling)

The objective of the master problem is to minimize the local generation cost (1) subject to operation constraints (2)-(11).

$$\min_{P} \sum_{t} \sum_{i \in G} F(P_{it}), \quad (1)$$

$$P_i^{\min} I_{it} \leq P_{it} \leq P_i^{\max} I_{it} \qquad \forall i \in G, \forall t, \quad (2)$$

$$T_{it}^{on} \geq UT_i (I_{it} - I_{i(t-1)}) \qquad \forall i \in G, \forall t, \quad (3)$$

$$T_{it}^{\text{off}} \geq DT_i(I_{i(t-1)} - I_{it}) \qquad \forall i \in G, \forall t, \quad (4)$$

$$P_{it} \leq P_{it}^{\text{dch,max}} u_{it} - P_{it}^{\text{ch,min}} v_{it} \qquad \forall i \in S, \forall t, \quad (5)$$

$$P_{it} \geq P_{it}^{\text{dch,min}} u_{it} - P_{it}^{\text{ch,max}} v_{it} \qquad \forall i \in S, \forall t, \quad (6)$$

$$u_{it} + v_{it} \leq 1 \qquad \forall i \in S, \forall t, \quad (7)$$

$$T_{it}^{\text{ch}} \geq MC_i(u_{it} - u_{i(t-1)}) \qquad \forall i \in S, \forall t, \quad (8)$$

$$T_{it}^{\text{dch}} \geq MD_i(v_{it} - v_{i(t-1)}) \qquad \forall i \in S, \forall t, \quad (9)$$

$$D_{dt}^{\min} z_{dt} \leq D_{dt} \leq D_{dt}^{\max} z_{dt} \qquad \forall d \in D, \forall t, \quad (10)$$

$$T_d^{on} \geq MU_d(z_{dt} - z_{d(t-1)}) \qquad \forall d \in D, \forall t. \quad (11)$$

Constraint (2) represents the maximum and minimum generation capacity of dispatchable units, where binary variable $I$ represents the unit commitment state that would be one when the unit is turned on and zero otherwise. Constraints (3) and (4) define the minimum up and down time limits, i.e., the minimum number of hours that a dispatchable unit must remain on/off once it is turned on/off. The energy storage is subject to minimum and maximum charging and discharging limits (5)-(7), where binary variables $u$ and $v$, respectively, represent the energy storage discharging and charging state which would be one when energy storage is discharging/charging and would be zero otherwise. Minimum charging/discharging time limits of energy storage are further defined in (8) and (9). Adjustable loads are subject to the minimum and maximum rated powers (10) and the minimum operating time (11), where binary variable $z$ represents the adjustable load state, which would be one when load is consuming power and zero otherwise.

### B. Subproblem (Optimal Operation)

The objective function of the subproblem is to find the worst-case optimal operation of local resources while minimizing the cost of power purchase from utility (12). This objective is subject to operation constraints (17)-(29).

$$\max_{U} \min_{P} \sum_t \rho_t^M P_t^M + \alpha, \quad (12)$$

$$I_{it} = \hat{I}_{it} \qquad \lambda_{it} \qquad \forall i \in G, \forall t, \quad (13)$$

$$u_{it} = \hat{u}_{it} \qquad \mu_{it}^{\text{dch}} \qquad \forall i \in S, \forall t, \quad (14)$$

$$v_{it} = \hat{v}_{it} \qquad \mu_{it}^{\text{ch}} \qquad \forall i \in S, \forall t, \quad (15)$$

$$z_{dt} = \hat{z}_{dt} \qquad \pi_{dt} \qquad \forall d \in D, \forall t \quad (16)$$

$$\sum_i P_{it} + P_t^M = \sum_d D_{dt} \qquad \forall t, \quad (17)$$

$$-P^{M,\max} \leq P_t^M \leq P^{M,\max} \qquad \forall t, \quad (18)$$

$$P_i^{\min} I_{it} \leq P_{it} \leq P_i^{\max} I_{it} \qquad \forall i \in G, \forall t, \quad (19)$$

$$P_{it} - P_{i(t-1)} \leq UR_i \qquad \forall i \in G, \forall t, \quad (20)$$

$$P_{i(t-1)} - P_{it} \leq DR_i \qquad \forall i \in G, \forall t, \quad (21)$$

$$P_{it} \leq P_{it}^{\text{dch,max}} u_{it} - P_{it}^{\text{ch,min}} v_{it} \qquad \forall i \in S, \forall t, \quad (22)$$

$$P_{it} \geq P_{it}^{\text{dch,min}} u_{it} - P_{it}^{\text{ch,max}} v_{it} \qquad \forall i \in S, \forall t, \quad (23)$$

$$C_{it} = C_{i(t-1)} - (P_{it} u_{it} \tau / \eta_i) - P_{it} v_{it} \tau \qquad \forall i \in S, \forall t, \quad (24)$$

$$C_i^{\min} \leq C_{it} \leq C_i^{\max} \qquad \forall i \in S, \forall t, \quad (25)$$

$$D_{dt}^{\min} z_{dt} \leq D_{dt} \leq D_{dt}^{\max} z_{dt} \qquad \forall d \in D, \forall t, \quad (26)$$

$$\sum_{t \in [\alpha_d, \beta_d]} D_{dt} = E_d \qquad \forall d \in D, \quad (27)$$

$$-\Delta^u - \Delta_t \leq P_t^M - P_{(t-1)}^M \leq \Delta^u - \Delta_t \qquad \forall t, \quad (28)$$

$$\Delta_t = \sum_j P_{jt}^c - \sum_j P_{j(t-1)}^c \qquad \forall t. \quad (29)$$

The objective function is maximized over uncertain parameter (distributed solar generation) and minimized over primary variables (i.e., local DER and load schedule and dispatch as well as utility grid power exchange). Hence, a robust solution (i.e., worst-case) will be calculated to ensure that microgrid can capture solar generation uncertainty. The binary scheduling variables are determined in the master problem and used in the subproblem as given values (13)-(16). The load balance equation (17) ensures that the sum of local DERs and power exchange with the utility grid would be equal to total microgrid load. The utility grid power $P^M$ could be negative (exporting power to the utility grid), positive (importing power from the utility grid), or zero. This power is limited to the capacity of the line/substation connecting the microgrid to the utility grid (18). Dispatchable units are subject to capacity limits (19) and ramping limits (20)-(21). The energy storage power in charging and discharging modes is limited by respective limits (22)-(23). The stored energy at energy storage is calculated based on charged and discharged power as well as roundtrip efficiency (24), and is further limited to its capacity limits (25). Parameter $\tau$ represent the time period of charging/discharging, which, in this paper, is assumed to be one hour. The adjustable load power is limited to its minimum and maximum rated powers (26) and daily energy limit required to complete an operating cycle (27).

To capture the solar generation variability, the microgrid power exchange with the utility grid, i.e., $P^M$, should be matched with the solar generation. This is carried out by adding (28) to the model. $\Delta^u$ represents the amount of variability that the utility grid is willing to capture where the rest is handled by the microgrid. $\Delta_t$ represents solar generation variability (in terms of difference between generation in two consecutive hours) and is calculated in (29). Constraints (28) and (29) accurately reflect solar generation variability as a constraint on the microgrid power exchange, thus allow an efficient capture of variability as targeted in this paper. Constraints (13)-(16) help determine the respective dual multipliers ($\lambda_{it}$, $\mu_{it}^{\text{dch}}$, $\mu_{it}^{\text{ch}}$, $\pi_{dt}$) to be used in forming the optimality cut. If the optimal solution is not obtained, which is checked by comparing the proximity of a lower bound from the master problem and an upper bound from the subproblem, the optimality cut (30) will be generated and sent back to the master problem for revising the current schedule.

$$\alpha \geq \hat{w} + \sum_t [\sum_{i \in G} \lambda_{it}(I_{it} - \hat{I}_{it}) + \sum_{i \in S} \mu_{it}^{\text{dch}}(u_{it} - \hat{u}_{it}) + \sum_{i \in S} \mu_{it}^{\text{ch}}(v_{it} - \hat{v}_{it}) + \sum_{d \in D} \pi_{dt}(z_{dt} - \hat{z}_{dt})] \quad (30)$$

The iterative process continues until the convergence criterion is met and the solution is proven optimal. It is assumed in the proposed model that the master problem solution will always result in a feasible solution when applied to the subproblem. This issue is further checked via numerical simulations on test distribution networks. However, in order to further make sure that the obtained schedule will never result in solution infeasibility in the subproblem, a feasibility check subproblem can be added and solved right after the master problem to revise obtained schedules and further ensure a feasible optimal operation. This subproblem will follow the standard Benders decomposition framework. Practically, the amount of variability that the utility grid is willing/able to capture ($\Delta^u$) is selected based on the day-ahead net load forecasts and grid operator's desired system flexibility. If the utility grid capability for capturing uncertainty and variability is less than the required amount (which is obtained based on net load forecasts), the grid operator can utilize distributed resources, such as microgrids, to compensate the flexibility shortage. Furthermore, the grid operator can calculate $\Delta^u$ via a cost-benefit analysis, i.e., to upgrade the current infrastructure to capture distributed solar generation uncertainty and variability or to procure the flexibility of existing microgrids for capturing variabilities and in turn pay for their service. Thus, in order to have this cooperation between microgrid and the utility grid, a robust communication infrastructure should be available between the utility grid and the microgrid, as well as between the utility grid and distributed prosumers. Additionally, the microgrid controller needs to be upgraded to consider additional constraints associated with flexibility.

## IV. NUMERICAL EXAMPLES

The performance of the proposed model is analyzed with a microgrid consisting of four dispatchable units, two nondispatchable units, including wind and solar, one energy storage, and five adjustable loads. The characteristics of dispatchable units are listed in Table I as well as the wind and solar generation, and the microgrid fixed load are tabulated in Table III [20]. The characteristics of energy storage, adjustable loads and the hourly market price are borrowed from [20]. The data of aggregated load, solar generation, and the net load of the sample distribution feeder in this study are listed in Table IV. A capacity of 10 MW is assumed for the line between the microgrid and the utility grid. The proposed microgrid optimal scheduling problem is solved using CPLEX 12.6 by a high-performance computing server consisting of four 10-core Intel Xeon E7-4870 2.4 GHz processors. The computation time is about 2 s for each simulation. It is worthwhile to mention that that the proposed model is generic and can be applied to any microgrid size without loss of generality.

TABLE I
SPECIFICATION OF DISPATCHABLE GENERATION UNITS

| Unit No. | Cost Coefficient ($/MWh) | Min.-Max. capacity (MW) | Min. Up/Down time (h) | Ramp Up/Down rate (MW/h) |
|---|---|---|---|---|
| 1 | 27.7 | 1 - 5 | 3 | 2.5 |
| 2 | 39.1 | 1 - 5 | 3 | 2.5 |
| 3 | 61.3 | 0.8 - 3 | 1 | 3 |
| 4 | 65.6 | 0.8 - 3 | 1 | 3 |

TABLE II
GENERATION OF NONDISPATCHABLE UNITS & MICROGRID FIXED LOAD (MW)

| Time (h) | 1 | 2 | 3 | 4 | 5 | 6 | 7 | 8 | 9 | 10 | 11 | 12 |
|---|---|---|---|---|---|---|---|---|---|---|---|---|
| Wind | 0 | 0 | 0 | 0 | 0.63 | 0.8 | 0.62 | 0.71 | 0.68 | 0.35 | 0.62 | 0.36 |
| Solar | 0 | 0 | 0 | 0 | 0 | 0 | 0 | 0 | 0 | 0 | 0 | 0.75 |
| Load | 8.73 | 8.54 | 8.47 | 9.03 | 8.79 | 8.81 | 10.1 | 10.9 | 11.2 | 11.8 | 12.1 | 12.1 |
| Time (h) | 13 | 14 | 15 | 16 | 17 | 18 | 19 | 20 | 21 | 22 | 23 | 24 |
| Wind | 0.40 | 0.37 | 0 | 0 | 0.05 | 0.04 | 0 | 0 | 0.57 | 0.6 | 0 | 0 |
| Solar | 0.81 | 1.20 | 1.23 | 1.28 | 1.00 | 0.78 | 0.71 | 0.92 | 0 | 0 | 0 | 0 |
| Load | 13.9 | 15.3 | 15.4 | 15.7 | 16.1 | 16.1 | 15.6 | 15.2 | 14.00 | 13 | 9.82 | 9.45 |

TABLE III
AGGREGATED PROSUMERS SOLAR GENERATION, CONSUMPTION, AND THE NET LOAD IN THE DISTRIBUTION FEEDER (MW)

| Time (h) | 1 | 2 | 3 | 4 | 5 | 6 | 7 | 8 | 9 | 10 | 11 | 12 |
|---|---|---|---|---|---|---|---|---|---|---|---|---|
| Solar | 0 | 0 | 0 | 0 | 0 | 0 | 0 | 0 | 1.0 | 4.0 | 8.0 | 11.5 |
| Load | 13.5 | 12.5 | 11.8 | 11.7 | 12.1 | 12.5 | 12.8 | 14.0 | 14.6 | 15.2 | 16.0 | 17.0 |
| Net load | 13.5 | 12.5 | 11.8 | 11.7 | 12.1 | 12.5 | 12.8 | 14.0 | 13.6 | 11.2 | 8.0 | 5.5 |
| Time (h) | 13 | 14 | 15 | 16 | 17 | 18 | 19 | 20 | 21 | 22 | 23 | 24 |
| Solar | 14.0 | 14.2 | 14.0 | 12.4 | 11.0 | 6.0 | 2.7 | 0.8 | 0 | 0 | 0 | 0 |
| Load | 18.5 | 18.0 | 17.0 | 16.7 | 17.0 | 18.0 | 20.3 | 20.7 | 19.0 | 17.0 | 14.5 | 13.8 |
| Net load | 4.5 | 3.8 | 3.0 | 4.3 | 6.0 | 12.0 | 17.6 | 19.9 | 19.0 | 17.0 | 14.5 | 13.8 |

A ±20% forecast error in distribution feeder solar generation is considered, along with a 12-hour/day budget as a limitation on uncertainty. A maximum of 2 MW/h is considered as the desired variability for the utility grid power ($\Delta^u$), which means that the net load variations above this amount should be captured by the microgrid. The proposed microgrid optimal scheduling considering prevailing uncertainties is solved for a 24-hour scheduling horizon.

Consideration of the microgrid in the distribution feeder without any contribution in capturing solar generation variability would exacerbate the net load variability due to price-based scheduling of the microgrid. The microgrid operation cost is $11,262.8 in this case. Considering a 2 MW/h variability limit, however, guarantees that the solar generation variability less than 2 MW/h is captured by the utility grid and beyond this amount is captured by the microgrid based on the proposed model. The solar generation variability is captured in this case, however, it increases the microgrid operation cost by 7.6% to $12,126.3. That is, the variability is captured, but at the expense of an increased operation cost for the microgrid.

Fig. 2 depicts the distribution feeder net load with and without considering the limit for solar generation variability (with ±20% forecast error) as well as the microgrid power exchange with the utility grid. The obtained results show that without any contribution from the microgrid, the utility grid should capture the severe net load changes at noon and early evening hours, i.e. a maximum of 15.75 MW/h between hours 11 and 12 as well as a maximum of 7.6 MW/h between hours 17 and 18. After the microgrid contribution all hourly changes are limited to 2 MW/h, as was desired by the grid operator. As Fig. 2 demonstrates, the microgrid decreases its power import from the utility grid in early evening hours and even exports the power to the utility grid in less expensive hours 19 and 22 to contribute to variability mitigation.

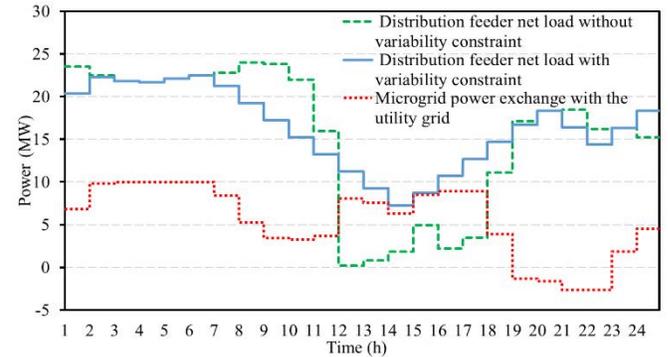

Figure 2. Distribution feeder net load, with and without variability constraint, and microgrid net load for uncertain solar generation.

The microgrid operation cost, however, is increased from $11,262.8 to $12,642.2. This 12.2% increase in the microgrid operation cost is the minimum that should be paid to the microgrid by the grid operator as an incentive for contribution in mitigating the solar generation variability. This cost difference can also be referred to as the microgrid lost opportunity in reaching its optimal schedule without limitations from the utility grid.

To further investigate the proposed model, the sensitivity of the microgrid operation cost with regards to the budget of uncertainty (hour per day) and the level of solar generation uncertainty (percentage of forecast error) is carried out and the results are tabulated in Table IV. The microgrid operation cost is increased for higher budgets of uncertainty and higher percentage of solar generation forecast error. The results show that increasing/decreasing the budget of uncertainty results in an increase/decrease in the microgrid operation costs, which, however, is not monotonically changing. For instance, by increasing solar forecast error from 5% to 25% microgrid operation cost increases 2.2% and 4.7% in the cases of 3 and 12 hours budget of uncertainty, respectively.

TABLE IV
MICROGRID OPERATION COST ($)

| Forecast Error (%) | Budget of Uncertainty (h) | | | | |
|---|---|---|---|---|---|
| | 0 | 3 | 6 | 9 | 12 |
| 5% | 12126.3 | 12187.8 | 12223.1 | 12229.7 | 12233.8 |
| 10% | 12126.3 | 12251.9 | 12328.7 | 12342.6 | 12356.4 |
| 15% | 12126.3 | 12319.4 | 12438.1 | 12458.8 | 12488.3 |
| 20% | 12126.3 | 12393.4 | 12561.9 | 12607.9 | 12642.2 |
| 25% | 12126.3 | 12465.1 | 12705.6 | 12790.8 | 12818.8 |

Fig. 3 depicts the microgrid operation cost for various amount of $\Delta^u$, in three cases of ignoring uncertainty and considering 10% and 20% forecast errors for distributed solar generation. Since with increasing $\Delta^u$ the microgrid should capture lower amounts of variability, the microgrid operation cost decreases. Furthermore, the obtained results show that the difference of the operation cost when ignoring and considering uncertainty significantly increases with decreasing the variability limit, while for larger limits these costs converge to the same values. It is mainly due to the fact that the variability limit in the microgrid optimal scheduling problem is relaxed and the solar generation variability is handled by the utility grid.

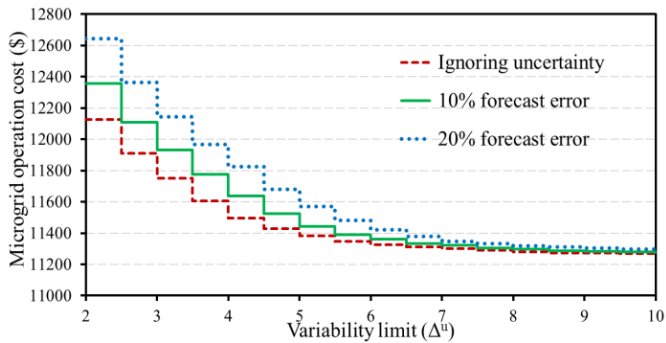

Figure 3. Microgrid operation cost for various variability limits, in three cases (ignoring uncertainty, 10% and 20% solar generation forecast error).

A larger test system is further studied to validate the applicability of the proposed model to various problem sizes. The considered test system includes twelve dispatchable units, fifteen adjustable loads, and three times larger aggregated distributed prosumers compared to the previous case. The results show the effectiveness of the proposed model in capturing solar generation uncertainty and variability (as variations are confined to 2MW/h), however the microgrid operation cost is increased due to providing the power to a larger aggregate load and capturing higher amounts of uncertainty and variability. The new operation cost, with 20% forecast error and budget of uncertainty of 12, is obtained as $37926.4, while the computation time is slightly increased (less than 3 seconds).

V. CONCLUSIONS

In this paper, a microgrid optimal scheduling model was proposed to capture solar generation variability and uncertainty, associated with variable nature of the solar power and the inherent uncertainty in solar power forecasts. A robust optimization method was proposed to find the worst-case microgrid optimal operation under uncertainty. The obtained results demonstrated the effectiveness of the microgrid application in capturing the solar generation variability and uncertainty. The numerical simulations further exhibited that by increasing the budget of uncertainty and percentage of forecast error, the microgrid operation cost increased as it was required to capture the higher amounts of variability and uncertainty on distributed solar generation.